\documentclass{ws-ijbc}
\usepackage{ws-rotating}     
\usepackage{graphicx}
\usepackage{epstopdf}
\usepackage{enumitem}
\usepackage{amsmath,amssymb}
\usepackage{amsfonts}
\usepackage{color}
\usepackage{mathtools}
\begin{document}



\title{GRAPHICAL STRUCTURE OF ATTRACTION BASINS OF HIDDEN CHAOTIC ATTRACTORS: THE RABINOVICH-FABRIKANT SYSTEM}

\author{MARIUS-F. DANCA}
\address{Romanian Institute of Science and Technology, \\400487 Cluj-Napoca, Romania\\     \email{danca@rist.ro}}

\author{PAUL BOURKE}
\address{The University of Western Australia, \\Crawley, Western Australia, 6009 Australia\\
     \email{paul.bourke@uwa.edu.au
}}

\author{NIKOLAY KUZNETSOV}
\address{Saint-Petersburg State University, St. Petersburg, Russia;\\
     Department of Mathematical Information Technology, \\
     University of Jyv\"{a}skyl\"{a}, Jyv\"{a}skyl\"{a}, Finland \\
     \email{nkuznetsov239@gmail.com}}

\maketitle
\begin{history}
\received{(to be inserted by publisher)}
\end{history}

\begin{abstract}
For systems with hidden attractors and unstable
equilibria, the property that hidden attractors are not connected with
unstable equilibria is now accepted as one of their main
characteristics. To the best of our knowledge
this property has not been explored using realtime interactive
three-dimensions graphics.
Aided by
advanced computer graphic analysis, in this paper, we explore this
characteristic of a particular nonlinear system with very rich and
unusual dynamics, the Rabinovich-Fabrikant system. It is shown that
there exists a neighborhood of one of the unstable equilibria within
which the initial conditions do not lead to the considered hidden
chaotic attractor, but to one of the stable equilibria or are
divergent. The trajectories starting from any neighborhood of the
other unstable equlibria are attracted either by the stable
equilibria, or are divergent.

\end{abstract}

\section{\label{sec:intro} Introduction}
One of the key tasks of the investigation of \emph{dynamical systems}
is the study of localization and analysis of \emph{attractors},
i.e.  the limited sets of the system states,
which are reached by the system from close initial data after transient processes.
While trivial attractors, i.e. stable equilibrium points,
can be easily revealed analytically or numerically,
the search of oscillating periodic or chaotic attractors
can turn out to be a challenging problem.

It is easy to visualize and describe the attractor
if there is only a single attractor
and its basin of attraction is the whole phase space.
However the situation becomes complicated
when the system is multistable with coexisting attractors \cite{PisarchikF-2014},
and some of the attractors are \emph{rare attractors}
with narrow basins of attraction \cite{ZakrzhevskySY-2007,DudkowskiJKKLP-2016}.
Various methods for the study of properties of the basin of attraction have been discussed,
e.g., in \cite{BrzeskiWKKP-2017}.

For numerical localization of an attractor
one needs to choose an initial point in the basin of attraction and observe how the trajectory, starting from this initial point, after a transient process reveals the attractor.
Computational errors, caused by a finite precision arithmetic and numerical integration
of differential equations, and sensitivity to initial data
allow one to get a representative visualization of the \emph{chaotic attractor}
by one pseudo-trajectory computed for a sufficiently large time interval.
Thus, from a computational point of view, it is natural
to suggest the following classification of attractors,
based on the simplicity of finding the basins of attraction in the phase space.
\emph{Self-excited attractors} can be revealed numerically by the integration of trajectories,
started in small neighborhoods of unstable equilibria,
while \emph{hidden attractors} have the basins of attraction
which are not connected with equilibria and are ``hidden somewhere'' in the phase space
\cite{LeonovKV-2011-PLA,LeonovK-2013-IJBC,KuznetsovL-2014-IFACWC,LeonovKM-2015-EPJST,Kuznetsov-2016}.

It should be noted that in the numerical computation of a trajectory over a finite-time interval
it is difficult to distinguish a hidden attractor from
a \emph{hidden transient chaos}
(a transient chaotic set in the phase space,
which can persist for a long time) \cite{DancaK-2017-CSF,ChenKLM-2017-IJBC,KuznetsovLMPS-2018}.\footnote{Note that generally the transient could have an
extremely long lifetime (superpersistent) \cite{greb}. On the other hand, if the time intervals are too large it could lead to
inaccurate numerical solutions.}


Hidden attractors are attractors in the systems without equilibria (see, e.g. rotating electromechanical systems with Sommerfeld effect (1902)
\cite{Sommerfeld-1902,KiselevaKL-2016-IFAC}), and in the systems with only one stable equilibrium (see, e.g. counterexamples \cite{LeonovK-2011-DAN,LeonovK-2013-IJBC}
to Aizerman's (1949) and Kalman's (1957) conjectures on the monostability of  nonlinear control systems \cite{Aizerman-1949,Kalman-1957}).

One of the first related problems is the second part
of 16th Hilbert problem \cite{Hilbert-1901}
on the number and mutual disposition of limit cycles
in two dimensional polynomial systems,
where nested limit cycles (a special case of multistability and coexistence of periodic attractors)
exhibit hidden periodic attractors (see, e.g., \cite{Bautin-1939,KuznetsovKL-2013-DEDS,LeonovK-2013-IJBC}).
The \emph{classification of attractors as being hidden or self-excited}
was introduced in connection with the discovery of the first hidden Chua attractor
\cite{KuznetsovLV-2010-IFAC,LeonovKV-2011-PLA,BraginVKL-2011,LeonovKV-2012-PhysD,KuznetsovKLV-2013,KiselevaKKKLYY-2017,StankevichKLC-2017}
and has captured much attention of scientists from around the world
(see, e.g. \cite{BurkinK-2014-HA,LiSprott-2014-HA,LiZY-2014-HA,PhamRFF-2014-HA,ChenLYBXW-2015-HA,KuznetsovKMS-2015-HA,SahaSRC-2015-HA,SemenovKASVA-2015,SharmaSPKL-2015-EPJST,ZhusubaliyevMCM-2015-HA,WeiYZY-2015-HA,DancaKC-2016,JafariPGMK-2016-HA,MenacerLC-2016-HA,OjoniyiA-2016-HA,PhamVJVK-2016-HA,RochaM-2016-HA,WeiPKW-2016-HA,Zelinka-2016-HA,BorahR-2017-HA,BrzeskiWKKP-2017,FengP-2017-HA,JiangLWZ-2016-HA,KuznetsovLYY-2017-CNSNS,MaWJZH-2017,MessiasR-2017-HA,SinghR-2017-HA,VolosPZMV-2017-HA,WeiMSAZ-2017-HA,ZhangWWM-2017-HA}).

For a \emph{self-excited attractor}
its basin of attraction
is connected with an unstable equilibrium
and, therefore, self-excited attractors
can be found numerically by the
\emph{standard computational procedure}
in which after a transient process a trajectory,
starting in a neighborhood of an unstable equilibrium,
is attracted to the state of oscillation and then traces it;
then the computations are performed for a grid of points
in the vicinity of the state of oscillation
to explore the basin of attraction and improve the visualization of the attractor.
Thus, self-excited attractors can be easily visualized
(e.g. the classical Lorenz and H\'{e}non  attractors
are self-excited with respect to all existing equilibria
and can be easily visualized by a trajectory from their vicinities).

For a hidden attractor, its basin of attraction is not connected with equilibria
and thus the search and visualization of hidden attractors
in the phase space may be a challenging task.

In this paper, using advanced graphical tools, we study the neighborhoods of all unstable equilibria in the case of the Rabinovich-Fabrikant system which has five equilibria. For the considered set of parameters, two equilibria are stable, while the other three are unstable. By visualizing the unstable equilibria, one can see that the property of the hidden attractor not being connected with unstable equilibria is confirmed. Also, the structure of the attraction basin of the hidden chaotic attractor is unveiled.

\section{Rabinovich-Fabrikant system}
In 1979, Rabinovich and Fabrikant introduced and analyzed from a physical point of view a model of the stochasticity arising from the modulation instability in a non-equilibrium dissipative medium. It is a simplification of a complex nonlinear parabolic equation modelling different physical systems: for the Tollmien-Schlichting waves in hydrodynamic flows, wind waves on water, concentration waves during chemical reactions in a medium in which diffusion occur, Langmuir waves in a plasma etc \cite{raba}.

The mathematical model of the Rabinovich-Fabrikant (RF) system is described by the following equations

\begin{equation}
\label{rf}
\begin{array}{l}
\overset{.}{x}_{1}=x_{2}\left( x_{3}-1+x_{1}^{2}\right) +ax_{1}, \\
\overset{.}{x}_{2}=x_{1}\left( 3x_{3}+1-x_{1}^{2}\right) +ax_{2}, \\
\overset{.}{x}_{3}=-2x_{3}\left( b+x_{1}x_{2}\right),
\end{array}%
\end{equation}

\noindent where $a,b>0$\footnote{Negative values also generate interesting dynamics, but this case does not have physical
meaning.}. The nature of the system dynamics is more sensitive to the parameter $b$ than to $a$, as such, $b$ is typically considered as the bifurcation parameter

For $a<b$ we have

\[
\textrm{div}(f(x))=\sum_{i=1}^{3}\frac{\partial }{\partial x_{i}}f_{i}(x)=2(a-b)<0.
\]

\noindent Therefore, the RF system is dissipative in the sense that the flow contracts in volume along trajectories, however it is not dissipative in the sense of Levinson \cite{LeonovKM-2015-EPJST}, i.e. there is no global bounded convex absorbing set. For example: the Lorenz system has a global bounded convex absorbing set and a global attractor; the classical Henon map has constant negative divergence but is not dissipative in the sense of Levinson and, thus, may have coexisting unbounded trajectories and local attractors with bounded basins of attraction \cite{jjj}; the system $\ddot x_1+x_1=0, \dot x_2 = -x_2$ has constant negative divergence $=-1$ but does not have bounded local attractors. Since in the RF system unbounded trajectories can be revealed numerically and there is no global attractor, the search of local hidden attractors is a challenging task.

The system is equivariant with respect to the symmetry

\begin{equation}\label{sim}
T(x_1,x_2,x_3)\rightarrow(-x_1,-x_2,x_3).
\end{equation}

\noindent This symmetry means that any orbit, has its symmetrical ''twin'' orbit in the sense that all orbits are symmetric one to another with respect to the $x_3$-axis.

Detailed numerical investigations, obtained by taking advantage of increasing computing resources, have been performed for the first time in \cite{danca1}, and despite the complicated form of the ODEs modeling the system, the RF system has been further numerically studied revealing new interesting aspects. Thus, beside transient chaos, numerically identified heteroclinic orbits\footnote{Because a complete mathematical analysis to reveal the existence and convergence of heteroclinic is highly problematic, the heteroclinic orbits have been found numerically aided by computer graphic analysis \cite{dancaxx}.} and several chaotic attractors with different shapes differentiated by energy-based analysis \cite{danca33}, the system exhibits not self-excited chaotic attractors and hidden chaotic attractors but, also chaotic transients \cite{dancaxx} and hidden chaotic transients \cite{dancay}.
 Also the system present multistability \cite{dancaxx}, several different equilibrium states coexisting for a given set of system parameters.

The symmetry \eqref{sim} is also reflected in the expression of the five equilibrium system's points: $X_0^*(0,0,0)$ and other four points

\begin{equation*}
\begin{array}{l}
X_{1,2}^{\ast }\left( \mp \sqrt{\dfrac{bR_1+2b}{4b-3a}},\pm \sqrt{b\dfrac{4b-3a}{R_1+2}},\dfrac{aR_1+R_2}{\left(4b-3a\right) R_1+8b-6a}\right),
\\
X_{3,4}^{\ast }\left( \mp \sqrt{\dfrac{bR_1-2b}{3a-4b}},\pm \sqrt{b\dfrac{4b-3a}{2-R_1}},\dfrac{aR_1-R_2}{\left(
4b-3a\right) R_1-8b+6a}\right) ,
\end{array}%
\end{equation*}

\noindent where $R_1=\sqrt{3a^{2}-4ab+4}$ and $R_2=4ab^{2}-7a^{2}b+3a^{3}+2a$.

\begin{figure}[t!]
\begin{center}
\includegraphics[clip,width=0.5\textwidth]{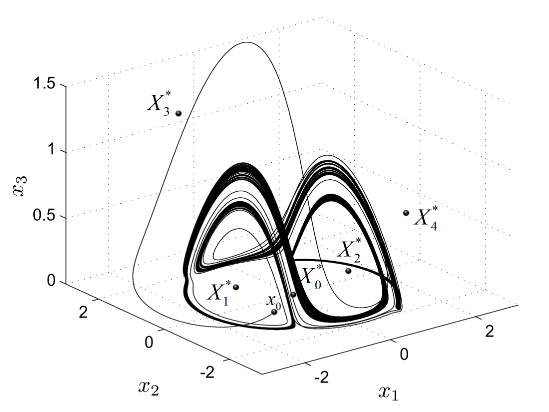}
\caption{Hidden chaotic attractor of the system RF \eqref{rf} for $a=0.1$ and $b=0.2876$. $X_{1,2}^*$ are stable equilibria, while $X_{0,3,4}$ are unstable equilibria.}
\label{figura1}
\end{center}
\end{figure}

One of the two hidden chaotic attractors, $H$ \cite{danca33}, is obtained for $a=0.1$ and $b=0.2876$ (Fig. \ref{figura1}).

A relatively large view (with the opposite corners $(-4,-4,0)$, $(0,4,8)$) of the attraction basins of the stable equilibria $X_{1,2}^*$, sliced with the plane $x_1=0$, is presented in Fig. \ref{figura7} (a) and (b) (blue representing the attractions basin points of the equilibrium $X_1^*$, while green the equilibrium $X_2^*$). The attraction basin of the hidden chaotic attractor $H$ is presented in Fig. \ref{figura7} (a), (c) and (d). In all images, grey is used to represent the points which generate divergence.

By having five equilibria, the RF system is topological non-equivalent to the classical systems, such as Lorenz and Chen systems (with three equilibria), R\"{o}ssler system (with two equilibria), some Sprott systems with one equilibrium \cite{spr} and so on.

As shown in \cite{dancaxx}, the equilibrium $X_0^*$ is an unstable equilibrium for every $a,b>0$. Furthermore, the line $x_1=x_2=0$, i.e. the $x_3$-axis, is invariant with
the reduced equation $\dot{x}_3=-2bx_3$, which has the solution $x_3(t)=e^{-2bt}x_3(0)$. Therefore, all points along $x_3$, which is one-dimensional stable manifold, will be attracted by $X_0^*$, but repulsed along spirals on the two-dimensional unstable manifold $x_3=0$.

\begin{figure}[t!]
\begin{center}
\includegraphics[clip,width=0.9\textwidth]{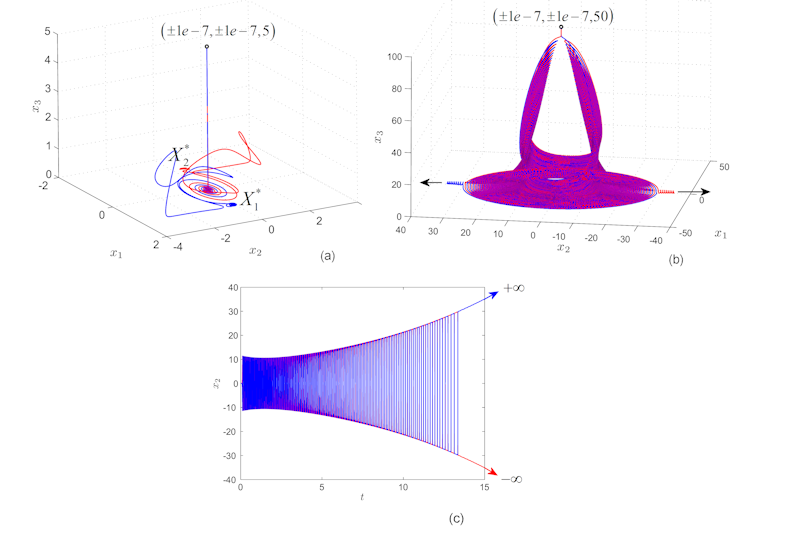}
\caption{Superimposed trajectories for the same parameters $a=0.1$ and $b=0.2876$, but for different initial conditions. (a) Initial conditions $x_0=(1e-7,1e-7,5)$ and $x_0=(-1e-7,-1e-7,5)$; (b) Initial conditions $x_0=(1e-7,1e-7,50)$ and $x_0=(-1e-7,-1e-7,50)$. (c) Time series ($x_2$) shows that there exist unstable transient oscillations, before the trajectories diverge.}
\label{figura2}
\end{center}
\end{figure}

To unveil one of the unusual dynamics of this system, for $a=0.1$ and $b=0.2876$, consider initial conditions close to $x_3$ axis (at distance smaller than $0.08$ (see Section 3), trajectories will tend, as mentioned before, to one of the stable equilibria $X_{1,2}^*$ (see blue trajectories to $X_1^*$ and red trajectories to $X_2^*$ in Fig. \ref{figura2} (a) for initial conditions $x_0=(1e-7,1e-7,5)$ and $x_0=(-1e-7,-1e-7,5)$ respectively). This happens, because points situated at distance smaller than about $0.08$ belong to one of attraction basins of $X_{1,2}^*$. However, for relative larger values $x_3(0)$ and $x_{1,2}(0)$ but still close to the axis $x_3$, trajectories are no longer attracted by $X_{1,2}^*$ but, after unstable transient oscillations tend to $\pm \infty$ along directions parallel to the axis $x_2$ (see Fig.  \ref{figura2} (b) where $x_0=(1e-7,1e-7,50)$ and $x_0=(-1e-7,-1e-7,50)$ respectively). Similar dynamics appear after some chaotic transient oscillations, which last for a relatively short time integration of about $t\in[0,20]$, for initial conditions on the plane $x_3=0$. Also, there exist divergent trajectories which are first attracted along directions parallel to the axis $x_2$, by some``virtual'' saddles, situated at large distances from unstable equilibria $X_{3,4}^*$), this kind of virtual points being presented in \cite{dancaxx}.

For the considered parameters, $a=0.1$ and $b=0.2876$, the equilibria $X_{1,2}^*$ are stable, while $X_0^*$ and $X_{3,4}$ are unstable.

Note that the considered hidden chaotic attractor $H$, coexists with three unstable equilibria \cite{danca33} (see also Fig. \ref{figura1}).

\section{Graphic analysis}
As outlined in the Introduction, an attractor is called a hidden attractor if its basin of attraction does not intersect with small neighborhoods of equilibria. Therefore, to verify that a chaotic attractor of RF's system is hidden, we might numerically check the trajectories starting from initial points situated in small vicinities of unstable equilibria. If there exist such neighborhoods whose points are either attracted by the stable equilibria or are divergent and do not lead to the attractor, then the attractor is considered as being hidden.

Generally, hidden chaotic attractors are found by searching thousands or tens of thousands of initial conditions in the neighborhoods of unstable equilibria, to see whether or not these neighborhoods lead to hidden attractors. This paper presents effective graphical visualizations that are both 3D and interactive of these neighborhoods and of the attraction basins of the one of the hidden chaotic attractors in the case of the RF system.

Consider the hidden attractor $H$ presented in Fig. \ref{figura1}.
In order to explore the attraction basin of $H$, parallelepipedic neighborhoods of unstable points have been considered. The points within these neighborhoods will be represented by small spheres. The intensive simulations, lead to a compromise between the computer time and accuracy of the graphical results. Thus, the fixed step of the parallelepiped scanning along the three axes is $1e-3$, or smaller and the size of the sphere radius is scaled so that nearby points (consecutive points, from the scanning point of view) are tangent spheres.

Exploring and understanding the nature of the basin of attraction
given the large number of three-dimensional points resulting from the
simulations is challenging. One goal was to be able to create
representations that could be studied interactively 3D rather than only
from fixed viewpoints or only in 2D sections. Another desirable, and often conflicting aim
was to subsequently be able to create high quality graphical
representations for publication. These goals were met by separating
the visualization software from the simulation software. Customized
tools were developed for the data visualization aspects of the
research, this software creates simplified geometry for realtime
exploration and at the same time high quality geometry for ray-traced
style rendering. In both cases points were represented by small glyphs,
typically spheres or boxes centered on each simulation point and
scaled to typically touch the neighboring points. Each point is
classified by the attractor it belongs to, this allows the various
basins of attraction to be shown individually or all together.

\vspace{3mm}
\begin{remark}
Different dedicated numerical methods for ODEs, implemented in different software packages, might give different results for the same parameters values and initial conditions. In this paper we utilized the Matlab integrator ode45 with $RelTol=1e-6$, and $AbsTol=1e-9$\footnote{In \cite{bren} the following relative tolerance is proposed: $RelTol= 10^{(m+1)}$, where $m$ is the precise number of digits.}  and the integration time $t\in[0,400]$. In order to have reliable numerical simulations, one has to find a compromise between the relative short time intervals where precise solutions can be found numerically (see e.g. \cite{sara,wang}) and the minimum time intervals required to reveal the nature of the systems, and avoid if necessary, the potential chaotic transients. Therefore, the results obtained in this paper are strongly related to the choice and setting up of the numerical method used to integrate this particular system.
\end{remark}

\begin{figure}[t!]
\begin{center}
\includegraphics[clip,width=0.8\textwidth]{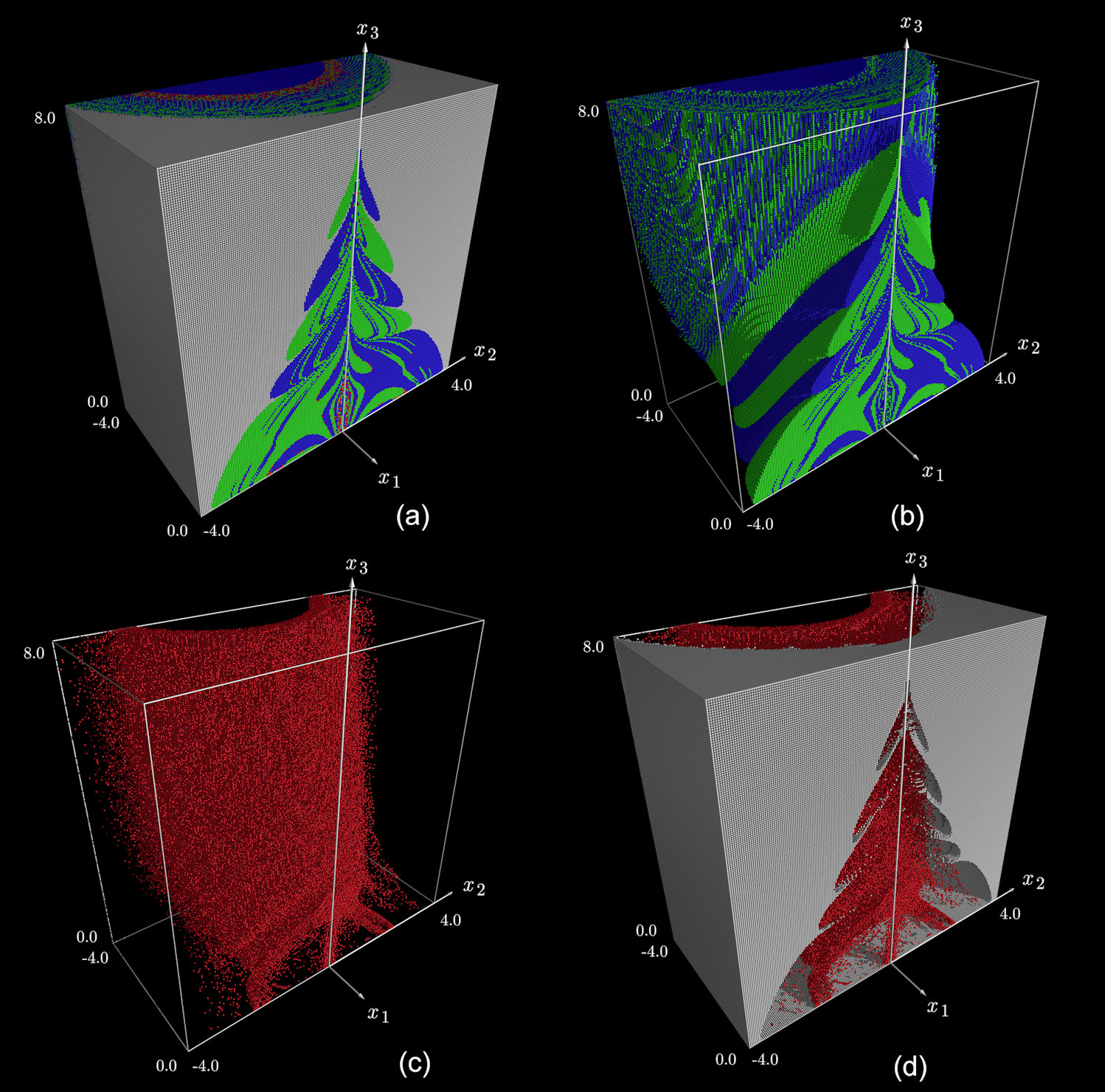}
\caption{(a) A relatively large view of attraction basins. (b) Attraction basins of the stable equilibria $X_{1,2}^*$; (c) Attraction basin of the hidden chaotic attractor $H$; (d) Attraction basin of the hidden chaotic attractor $H$ together with divergence points.}
\label{figura7}
\end{center}
\end{figure}

\begin{figure}[t!]
\begin{center}
\includegraphics[clip,width=0.75\textwidth]{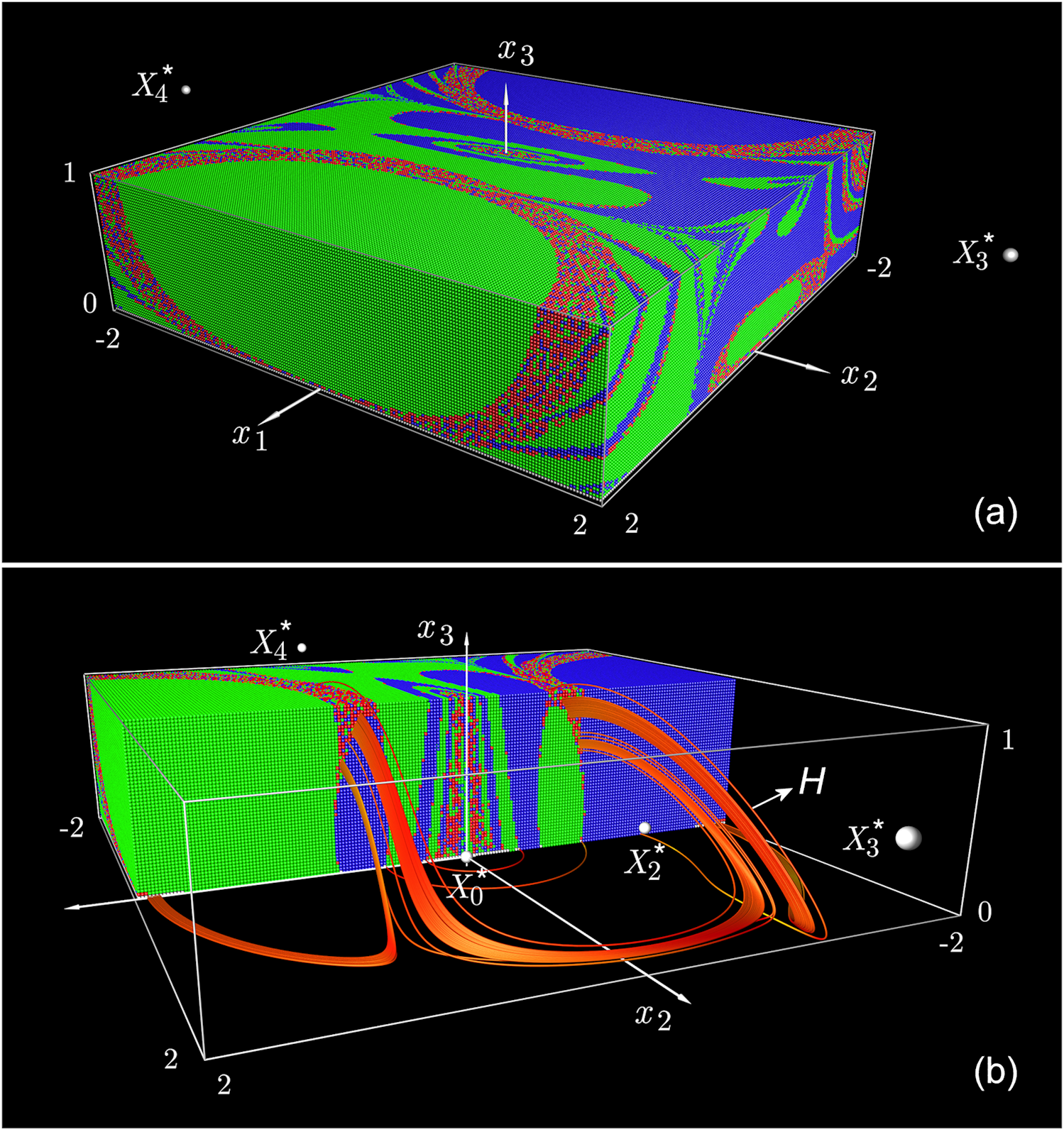}
\caption{Parallelepiped neighborhood of the unstable equilibrium $X_0^*$. Blue points represent the initial conditions which lead to the stable equilibrium $X_1^*$, green points lead to the stable equilibrium $X_2^*$, while the red points lead to the hidden attractor $H$; (a) General view; (b) Hidden attractor overplotted on the parallelipipedic neighborhood.}
\label{figura3}
\end{center}
\end{figure}

Let first consider the unstable equilibrium $X_0^*$. The explored domain, a zoomed parallelepipedic detail from Fig. \ref{figura7} (a), with opposite corners $(-2,-2,-2)$, $(2,2,1)$, contains $2,000,000$ points so that it includes the hidden attractor $H$ (see Fig.\ref{figura3} (a)). This large number of points determines a sufficiently accurate graphical analysis of the neighborhoods of unstable equilibria.

\begin{figure}[t!]
\begin{center}
\includegraphics[clip,width=0.9\textwidth]{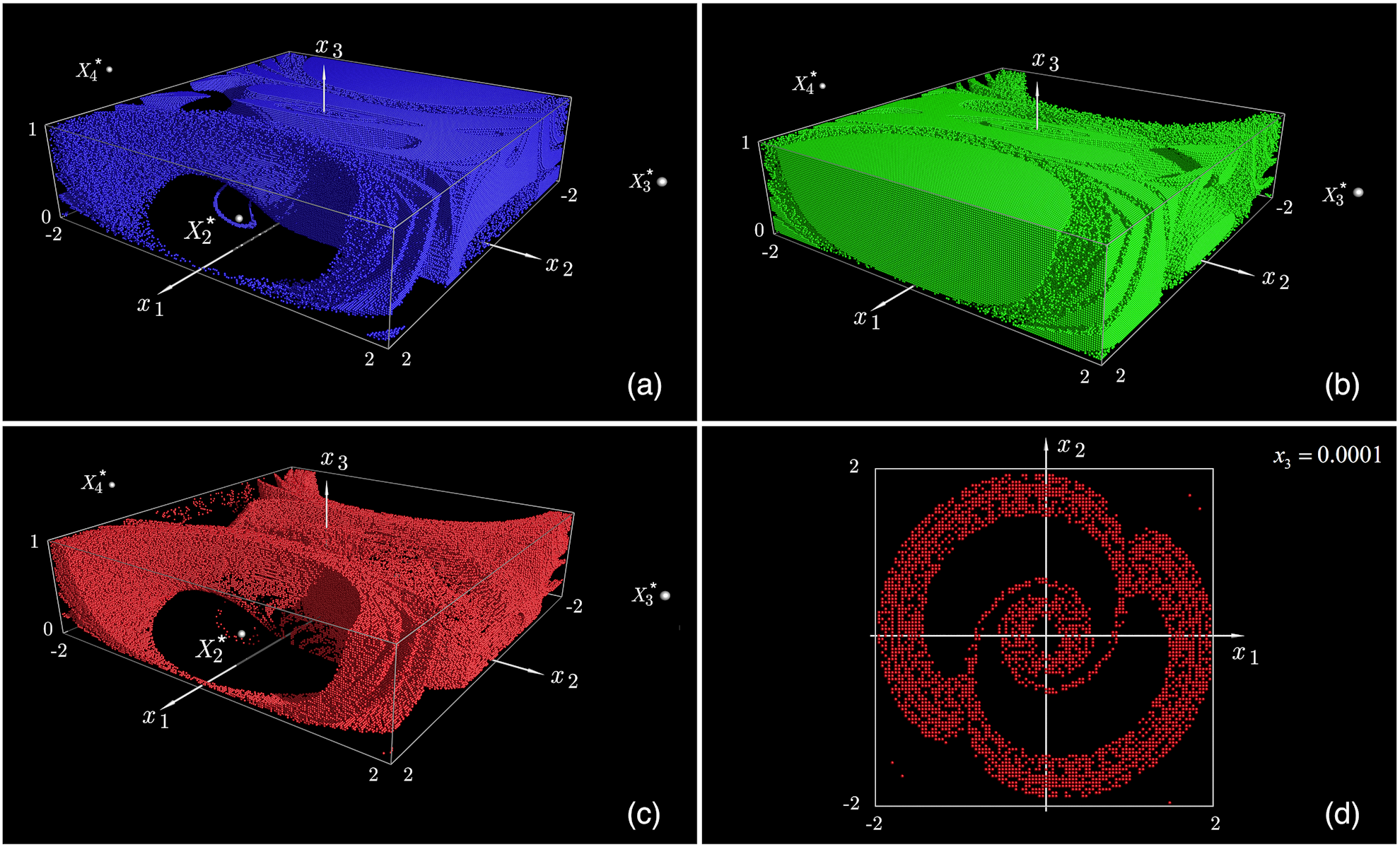}
\caption{Separate views of partial attraction basins; (a) Points of the attraction basins of the stable equilibrium $X_1^*$ (not visible here); Points of the attraction basins of the stable equilibrium $X_2^*$ (not visible here); (c) Points of the attraction basins of the hidden attractor $H$; (d) Section through the attraction basin in Fig. \ref{figura4} (c) with the plane $x_3=0.0001$.}
\label{figura4}
\end{center}
\end{figure}

As shown in Fig. \ref{figura7} (a), the points considered as initial points of the numerical integration of the system which lead to $X_1^*$ are colored blue (Fig. \ref{figura4} (a)), those leading to $X_2^*$ in green (Fig. \ref{figura4} (b)), and points of the attraction basin of the hidden attractor, in red (Fig. \ref{figura4} (c)). Divergency points (plane $x_3=0$) are not shown.

It can be seen that the hidden attractor $H$ evolves along the frontiers of attraction basins of the two stable equilibria (blue and green), where there are a mixed (possible a fractal set) of points attracted by $X_1^*$ and $X_2^*$ (see Fig. \ref{figura3} (b)). Moreover, these frontiers also contains initial points of $H$.

Even at first sight the origin seems to be surrounded by initial red points which lead to $H$, the ``closest'' horizontal section to the plane $x_3=0$ (Fig. \ref{figura4} (d)), unveils that one can find an empty disc-like region without red points, of radius of about $0.08$, in concordance with the property of hidden chaotic attractors. By the ``closest'' horizontal plane one understands here the first scanned plane which, in this case, is situated at a distance $x_3=0.0001$ from the plane $x_3=0$. One can suppose that this region exists also for $x_3\rightarrow 0$.

\begin{figure}[t!]
\begin{center}
\includegraphics[clip,width=0.8\textwidth]{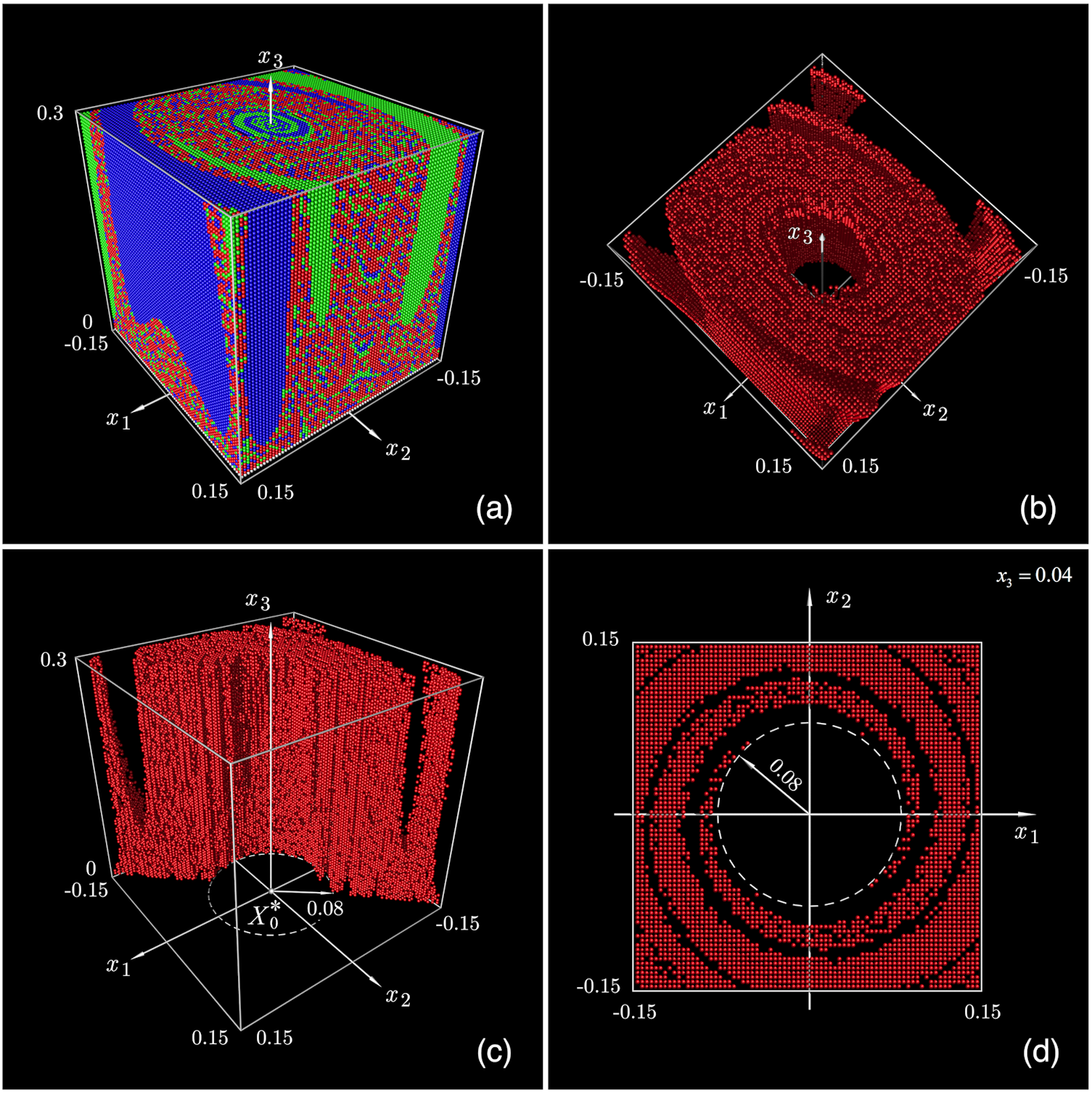}
\caption{Zoomed detail of the parallelipiped in Fig. \ref{figura3} (a), containing the unstable equilibrium $X_0^*$; (a) General view, revealing the possible fractal structure of the basins boundaries; (b) View of the initial points leading to the hidden attractor $H$ unveiling the empty space around the unstable point $X_0^*$; (c) Vertical section through the attraction basin of $H$ in Fig. \ref{figura5} (b) presenting the cylinder-like shape of the empty region, of ray $0.08$; (d) Horizontal section with the plane $x_3=0.04$.}
\label{figura5}
\end{center}
\end{figure}

 A zoomed three-dimensional detail of the neighborhood of unstable equilibrium $X_0^*$, beside the possible fractal structure (Fig. \ref{figura5} (a)), reveals the fact that the ``empty-red'' region persists along the $x_3$ axis, having a cylinder-like shape (Fig. \ref{figura5} (b) and (c)). In other words, all initial points within this ``empty-red'' region lead, by integration, to one of the stable equilibria, $X_1^*$ (blue points) or $X_2^*$ (green points), or are divergent (if the points are on the plane $x_3=0$) and not attracted by $H$. Another horizontal section with the plane $x_3=0.04$ is presented in Fig. \ref{figura5} (d) and, compared with that in Fig. \ref{figura4} (c), indicates that the shape of the boundary of the empty region changes slightly with the variations of $x_3$, but has the same radius size of about $0.08$ (Fig. \ref{figura5} (d)).

\begin{figure}[t!]
\begin{center}
\includegraphics[clip,width=0.8\textwidth]{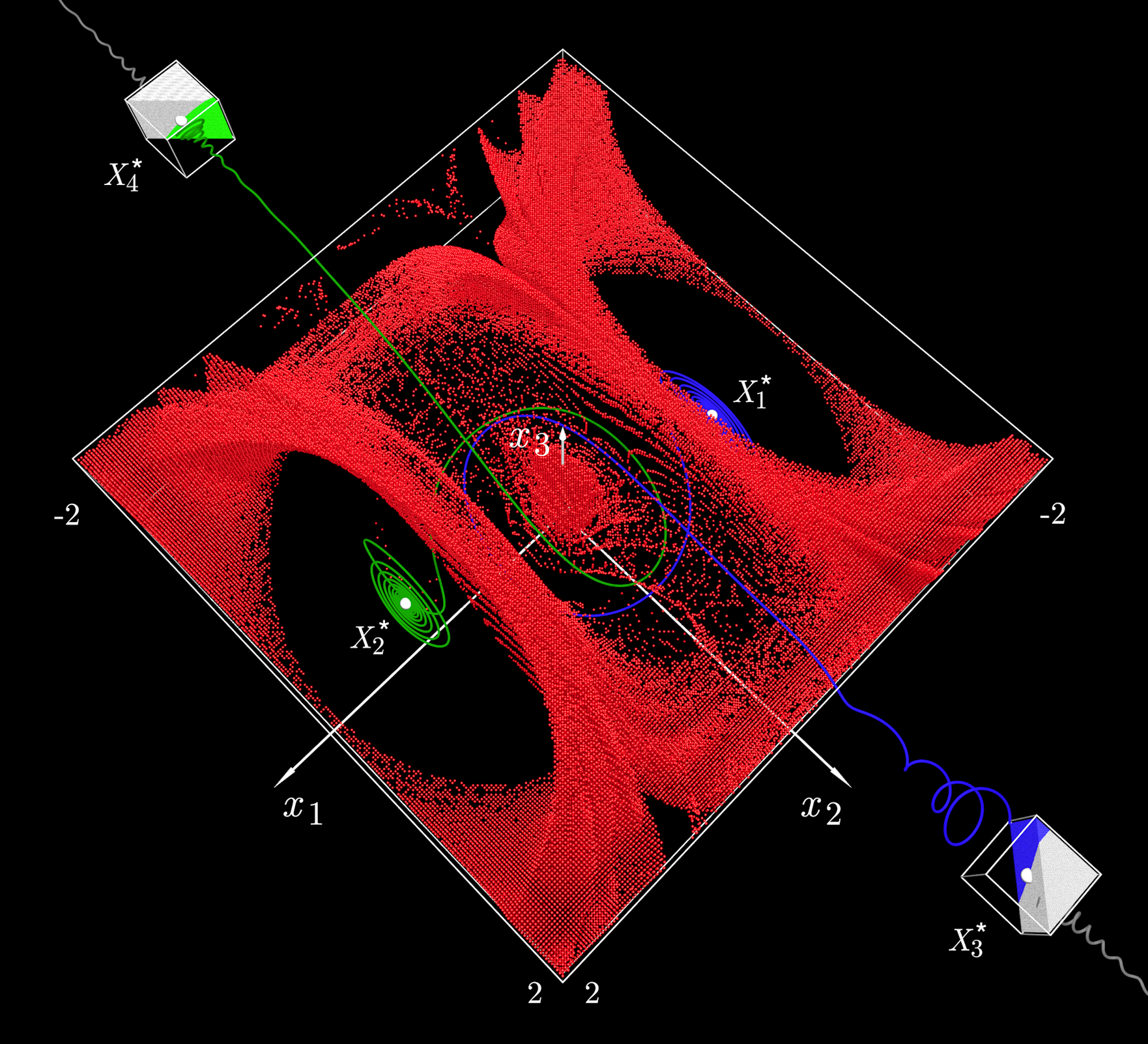}
\caption{General view of all equilibria and their partial attraction basins. From any points in the neighborhoods of the unstable equilibria $X_{3,4}^*$ the trajectories tend either to the stable equilibrium $X_1^*$ (blue), or to the stable equilibrium $X_2^*$ (green), or diverge (gray). Attraction basins of $X_{1,2}^*$ are not drawn here.}
\label{figura6}
\end{center}
\end{figure}

Consider now the unstable equilibria $X_{3,4}^*$, as shown in Fig. \ref{figura6}. As can be seen, these equilibria are not related to $H$. The trajectories starting from any neighborhood of these equilibria are attracted either by $X_1^*$ (blue points and trajectories) or by $X_2^*$ (green points an trajectories), or tend to infinity (grey points and trajectories).

Summarizing, all unstable equilibria of the RF system are not connected with $H$. In other words, $H$ cannot be found by starting from initial points of however small neighborhoods of unstable equilibria, such as for self-excited attractors and the largest neighborhood of $X^*_0$, which has no connection with the attraction basin of the hidden attractor $H$ has a well defined unbounded cylinder-like shape. Moreover, for this system the attraction basin of $H$ seems unbounded (see also Fig. \ref{figura7} (a)).


\section*{Conclusion}

In this
paper, for the case of the RF system and aided by advanced computer
graphic techniques, we visualized both the attraction basins of the
stable equilibria and the attraction basin of the considered hidden
chaotic attractor. We illustrated that the attraction basin of the
hidden chaotic attractor is not connected with unstable equilibria.
Also, it is shown that one of the unstable equilibria is connected with the attraction basin of the hidden chaotic attractor, only outside an unbounded cylinder-like neighborhood of the unstable equilibria.
\vspace{3mm}

\section*{Appendix}
\noindent Interactive versions of the 3D figures can be found online at the following address

https://sketchfab.com/pbourke/collections/hidden-attractors

\noindent Alternatively, 3D models in the OBJ format can be downloaded from here

http://paulbourke.net/papers/hidden-attractors

\noindent OBJ models can be opened in a wide range of 3D viewing software packages.

\section*{Acknowledgements} \label{sec:acknowledgement}
The work is done within  the Russian Science Foundation project (14-21-00041)

\end{document}